# Phase shifting Interferometry and Wavefront Compensation using Liquid Crystal on Silicon (LCOS) Spatial Light Modulator (SLM)

**Indrani Bhattacharya[1], Marcus Petz[2], Rainer Tutsch[3]**

[1]indrani.bhattacharya@tu-braunschweig.de, Visiting Scientist and Researcher, Institute of Production Metrology (IPROM), Techniche Universitat, Braunschweig, Germany.
[2]m.petz@tu-braunschweig.de Associate Professor, Institute of Production Metrology (IPROM), Techniche Universitat, Braunschweig, Germany.
[3]r.tutsch@tu-braunschweig.de Professor and Director, Institute of Production Metrology (IPROM), Techniche Universitat, Braunschweig, Germany (Corresponding Author).

**Abstract:** We present a simple and novel technique for interferometric surface measurement using a Liquid Crystal on Silicon (LCOS) Spatial Light Modulator (SLM) as phase shifter and wavefront-compensator simultaneously. In comparison to a mechanical piezo-based phase shifter measurement, speed and measurement range can be increased by this approach.

### Section 1: Introduction

**Phase Shifting Interferometry (PSI)** is a powerful tool widely used in optical metrology for high-precision surface profiling and wavefront measurement [1][2]. By introducing controlled phase shifts into an interferometric setup, PSI enables the retrieval of quantitative phase information with sub-wavelength accuracy [3]. Its sensitivity and precision make it indispensable in applications ranging from optical testing and surface metrology [4][5] to adaptive optics [6] and biomedical imaging [7].

Recent advances in programmable optical devices, particularly **Liquid Crystal on Silicon Spatial Light Modulators (LCOS SLMs)**, have significantly enhanced the flexibility and functionality of interferometric systems [8][9]. LCOS SLMs can dynamically modulate the phase of incident light with high spatial resolution, making them suitable for wavefront control and phase manipulation [10][11]. They are increasingly being employed as phase shifters in PSI setups, replacing traditional mechanical components to achieve electronically controlled phase shifts [12]. Their ability to generate arbitrary phase patterns with pixel-level precision enables precise and repeatable phase stepping without mechanical movement, thereby improving system stability, increasing speed, and facilitating real-time measurements [13][14][15][16][17].

This paper explores the application of LCOS SLM-based PSI for real-time wavefront compensation. Section 2 presents the theoretical background of PSI and the experimental arrangement is shown in Section 3. Section 4 describes the determination of Phase-shift to gray value response of Holoeye Pluto-2.1-VIS-016 LCOS display. Section 5 describes the best gray values for Four-step phase-shift measurement and identification of possible phase errors using Zernike polynomial. Section 6 describes the method for wavefront compensation. Section 7 refers to the conclusion and discussion and section 8 refers to the further scope of research.

### Section 2: Theoretical Background

Phase Shifting Interferometry (PSI) is a technique used to retrieve phase information from an interference pattern by introducing known phase shifts between successive interferograms. Among the various phase shifting schemes, the **four-step phase shifting algorithm** is one of the most widely used due to its balance of accuracy, computational simplicity, and noise robustness. In a typical PSI setup, the recorded interferogram intensity $I_k(x, y)$ at each pixel $(x, y)$ is given by:

# Phase shifting Interferometry and Wavefront Compensation using Liquid Crystal on Silicon (LCOS) Spatial Light Modulator (SLM)

**Indrani Bhattacharya[1], Marcus Petz[2], Rainer Tutsch[3]**

$$I_k(x,y) = I_0(x,y) + I_m(x,y)\cos[\phi(x,y) + \delta_k] \quad (1)$$

Where:

- $I_0(x,y)$ is the background or average intensity,
- $I_m(x,y)$ is the fringe modulation (amplitude),
- $\phi(x,y)$ is the phase to be determined,
- $\delta_k$ is the known phase shift for the $k$ th interferogram.

In the four-step algorithm, four interferograms are recorded with uniform phase shifts of δ=90°=π/2. The phase shifts are:

$$\delta_1 = 0, \ \delta_2 = \frac{\pi}{2}, \ \delta_3 = \pi, \ \delta_4 = \frac{3\pi}{2} \quad (2)$$

The corresponding intensity equations are:

$$I_1 = I_0 + I_m \cos[\phi(x,y)] \quad (3)$$

$$I_2 = I_0 + I_m \cos\left[\phi(x,y) + \frac{\pi}{2}\right] = I_0 - I_m \sin[\phi(x,y)] \quad (4)$$

$$I_3 = I_0 + I_m \cos[\phi(x,y) + \pi] = I_0 - I_m \cos[\phi(x,y)] \quad (5)$$

$$I_4 = I_0 + I_m \cos\left[\phi(x,y) + \frac{3\pi}{2}\right] = I_0 + I_m \sin[\phi(x,y)] \quad (6)$$

The unknown phase $\phi(x,y)$ can be determined by subtracting and combining Equations (3),(4),(5) and (6) as:

$$I_4 - I_2 = 2I_m \sin[\phi(x,y)] \quad (7)$$

$$I_1 - I_3 = 2I_m \cos[\phi(x,y)] \quad (8)$$

Dividing Equations (7) by (8),

$$\frac{I_4 - I_2}{I_1 - I_3} = \tan[\phi(x,y)] \quad (9)$$

Rearranging, we get the result for the four step PSI algorithm as:

$$\phi(x,y) = \tan^{-1}\left[\frac{I_4 - I_2}{I_1 - I_3}\right] \quad (10)$$

Equation (10) is evaluated at each measurement point to obtain the map of wrapped phase of the measured wavefront.

The raw phase values extracted from the interferograms are wrapped phases and they are confined to the interval (0, 2π) which happens due to the periodic nature of the arctangent function used in Equation (10). Using the four-step phase-shifting algorithm mentioned above, the expression for wrapped phase is determined using the mathematical expression ensuring that the result lies in the interval [0, 2π]:

$$\phi_{wrapped}(x,y) = mod\left[\left(\tan^{-1}\left[\frac{I_4 - I_2}{I_1 - I_3}\right]\right) + 2.\pi, \ 2.\pi\right] \quad (11)$$

Equation (11) determines the wrapped phase where discontinuity appears for a phase jump of 2π. The Phase unwrapping process is used to reconstruct the original, continuous phase $\phi(x,y)$ from the wrapped values by adding 2π or integral multiples of it [2][18]. To remove the discontinuities that are present in the raw phase data generated in the wrapped phase using 4-step PSI the following mathematical equation is used:

$$\phi_{unwrapped}(x,y) = \phi_{wrapped}(x,y) + 2\pi \cdot k(x,y) \qquad (12)$$

Where $k(x,y)$ is an integer chosen to ensure that the result is as smooth as possible.

The unwrapped phase $\phi_{unwrapped}(x,y)$ is proportional to the optical path difference (OPD). As the system we are using is a reflection interferometer, the OPD is twice the surface height $h(x,y)$ and for a wavelength λ, the relation is:

$$h(x,y) = \frac{\lambda}{4\pi} \cdot \phi_{unwrapped}(x,y) \qquad (13)$$

The height map or surface profile represents the topography of the surface we are measuring.

Zernike polynomials, $Z_n^m(r,\theta)$, a complete set of orthogonal polynomials defined over the unit disk (circle with radius 1), are used here to model and subtract low-order optical aberrations from the measured height map. The result is a residual surface that reflects fine shape variations after removing systematic/global trends. $Z_n^m(r,\theta)$ is expressed in the general form :

$$Z_n^m(r,\theta) = R_n^m(r) \cdot \begin{cases} \cos(m\theta), & m \geq 0 \\ \sin(-m\theta), & m < 0 \end{cases} \qquad (13)$$

Where:

- $n$ is the radial degree (non-negative integer),
- $m$ represents azimuthal frequency which is an integer and $|m| \leq n$, $n - m \ even$,
- $r \in (0,1), \ \theta \in (0,2\pi)$,
- $R_n^m(r)$ is the radial component defined by:

$$R_n^m(r) = \sum_{s=0}^{(n-m)/2} (-1)^s \cdot \frac{(n-s)!}{s!\left(\frac{n+m}{2}-s\right)!\left(\frac{n+m}{2}-s\right)!} \cdot r^{n-2s} \qquad (14)$$

After the generation of an unwrapped surface $\hat{S}(r,\theta)$, we have fitted it as a linear combination of Zernike polynomials given by:

$$\hat{S}(r,\theta) = \sum_{n=1}^{N} a_n \cdot Z_n^m(r,\theta) \qquad (15)$$

Where:

- $Z_n^m$ are the basis functions, here Zernike Polynomials,
- $a_n$ are the coefficients (to be determined),
- N is the number of Zernike terms up to a chosen order.

**Phase shifting Interferometry and Wavefront Compensation using Liquid Crystal on Silicon (LCOS) Spatial Light Modulator (SLM)**
**Indrani Bhattacharya[1], Marcus Petz[2], Rainer Tutsch[3]**

We solved this by using least-squares fitting problem by minimizing:

$$\min_{a} \ \|Za - S\|^2 \tag{16}$$

Where:

- $Z \in \Re^{M \times N}$ is the design matrix of Zernike basis functions,
- $S \in \Re^{M}$ is the vector of surface height or phase at M valid points,
- $a \in \Re^{N}$ is the coefficient vector.

The solution is

$$a = Z^T . \hat{S} \tag{17}$$

After fitting, the surface is reconstructed and subtracted using the mathematical identity:

$$\text{Residual}\,(r, \theta) = S(r, \theta) - \hat{S}(r, \theta) \tag{18}$$

This residual surface contains the non-Zernike content i.e., the detailed variations that don't match any Zernike polynomial up to the chosen order, which is 6 in our case. These residuals highlight errors like surface roughness, mid-spatial frequency errors and local manufacturing defects. Zernike fitting is done on the **unit circle** via normalization of pixel coordinates.

A fringe contrast measurement is done for each interferogram recorded to ensure the clarity of the fringes observed using the formula:

$$\gamma(x, y) = \frac{I_{max}(x,y) - I_{min}(x,y)}{I_{max}(x,y) + I_{min}(x,y)} \tag{19}$$

Gamma value is measured in the range of $\gamma \in [0,1]$ where:

- $\gamma \approx 1$ means High Contrast and Good Visibility
- $\gamma \approx 0$ means Low Contrast and Poor Visibility.

**Section 3: Experimental Arrangement**

The schematic diagram of the experimental arrangement of LCOS SLM Phase Shifting Interferometer with Twyman- Green configuration is shown in *Figure 1* where a normal incidence from a He-Ne laser beam of wavelength 632.8 nm is collimated and allowed to pass through a polarising filter. The polarising filter is adjusted in such a way that the incident polarization is kept along the long display axis of the LCOS SLM. The polarised light is passed through a neutral filter and incident on a microscope objective of 40x magnification. A pin-hole of a diameter of 10 μm is placed at the back focal plane of the microscope objective and the spatially filtered divergent beam is propagating to a collimating lens of focal length  50 mm. The collimated beam of light is incident on a Cubic Beam Splitter (CBS) from where the beam divides into two parts. The reflected part from the CBS is directed to the reference path where the LCOS SLM, Model - Holoeye PLUTO-2.1-VIS-016 (420nm-650nm) is placed with its surface normal to the beam of light. The transmitted part of the light from the CBS is incident on  a convergent lens followed by a convex spherical mirror with its centre of curvature adjusted at the focus of the lens. By shifting the convex mirror away from this position, a well-defined aspheric component can be superimposed to the reflected spherical wave.

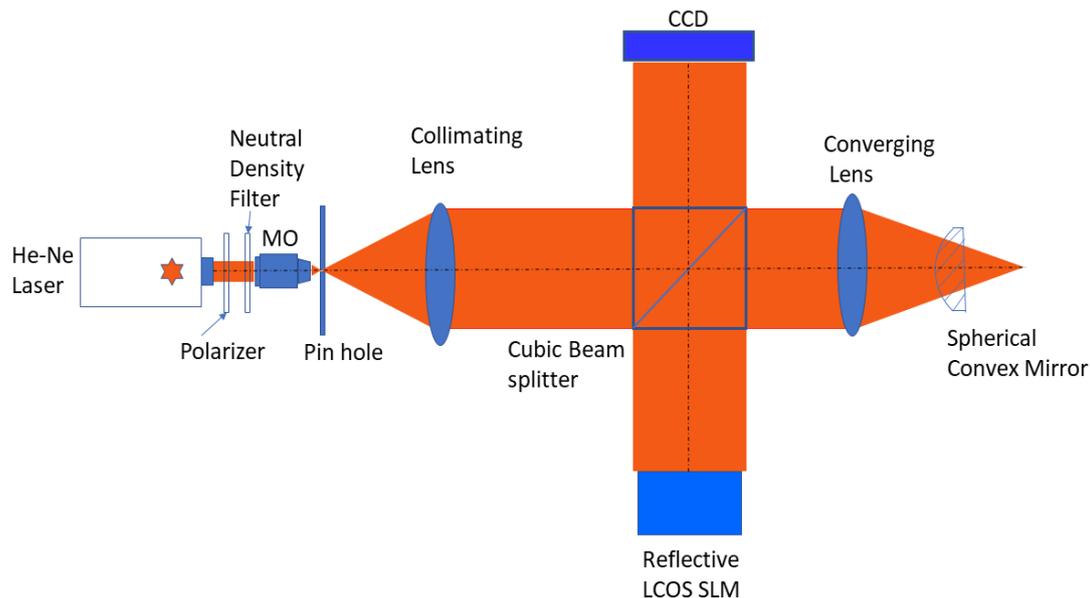

*Figure 1* The schematic diagram of the experimental arrangement of LCOS SLM Phase Shifting Interferometer with Twyman- Green configuration.

**Section 4: Determination of Phase-shift to gray value response**

In order to evaluate the characteristic curve of the LCOS display, the relation between the applied gray value and the resulting phase shift from the measurements of the global phase shift was performed using a Matlab script based on the approach by Sun et al. [19]. As the algorithm fails for phase shifts close to 0 and 2π, the phase measurement was repeated with two different reference values, the first being a gray value of 0 and the second being a gray value of 128. The valid responses from these two computations were then combined to obtain one complete measurement. The results obtained from two independent runs of the approach are shown in *Figure 2*. As can be seen, the response is relatively linear. The observed non-linearities appear to be mostly random, which results from environmental influences. It can also be seen that the maximum phase shift, reached (or expected) at a gray value of 255 seems to be slightly larger than 2π.

As presumably the calibration measurements performed by Holoeye are more stable than the measurements taken in the IPROM lab, the assumption was made that the LCOS delivers a sufficiently linear response (if driven with the calibration provided by Holoeye) but with a slightly larger range than 2π. In order to apply correct phase values, to perform a four-step phase shift measurement, the slope of the (linear) characteristic curve has to be determined. In order to be more robust against environmental influences the gray value that corresponds to a phase shift of 2π was computed from a small set of interferograms taken at gray values of 0, 120, 235, 240, and 245. From this it was found that a gray value of 240 delivers a phase shift close to 2π relative to a gray value of 0.

**Phase shifting Interferometry and Wavefront Compensation using Liquid Crystal on Silicon (LCOS) Spatial Light Modulator (SLM)**
**Indrani Bhattacharya[1], Marcus Petz[2], Rainer Tutsch[3]**

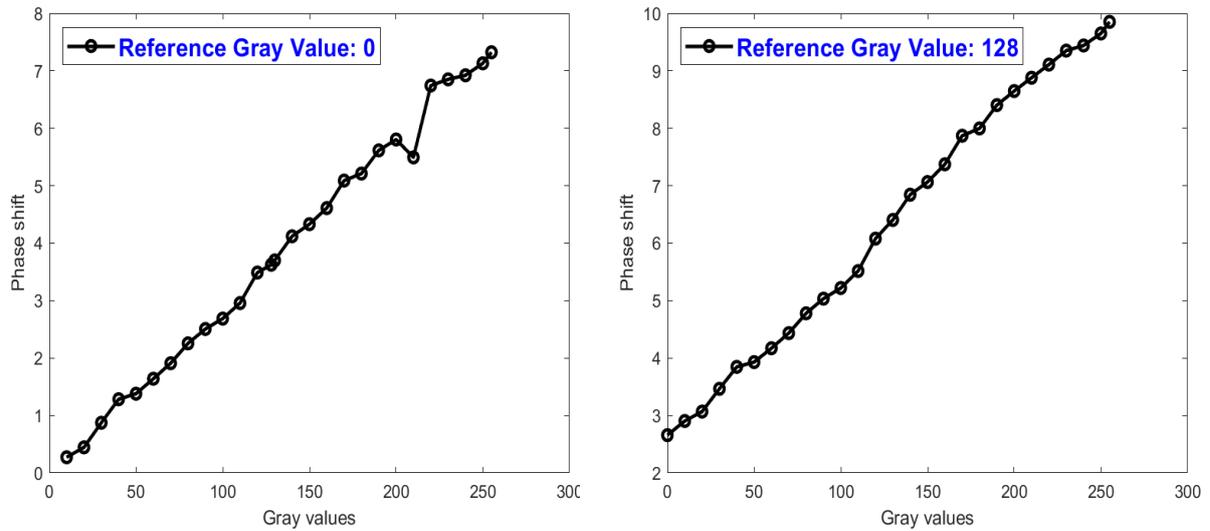

*Figure 2 LCOS measurement curves showing resultant Phase-shifts with the applied gray values for reference phases corresponding to Reference gray values 0 and 128.*

**Section 5: Four-step phase shift measurement**

From the above assumptions and findings, the best gray values for a four-step phase shift measurement were determined to be at 0, 60, 120, and 180. *Figure 3* shows exemplary interferograms on the left-hand side and the surface reconstructed from a four-step phase shift measurement using the above mentioned gray values on the right-hand side. For computation the original camera images were cropped to approximately the active area of the LCOS, as indicated by the red rectangular area in *Figure 3*. For unwrapping, an algorithm by Herraez et al. [20] in an Matlab implementation freely provided by Kasim [21] was used.

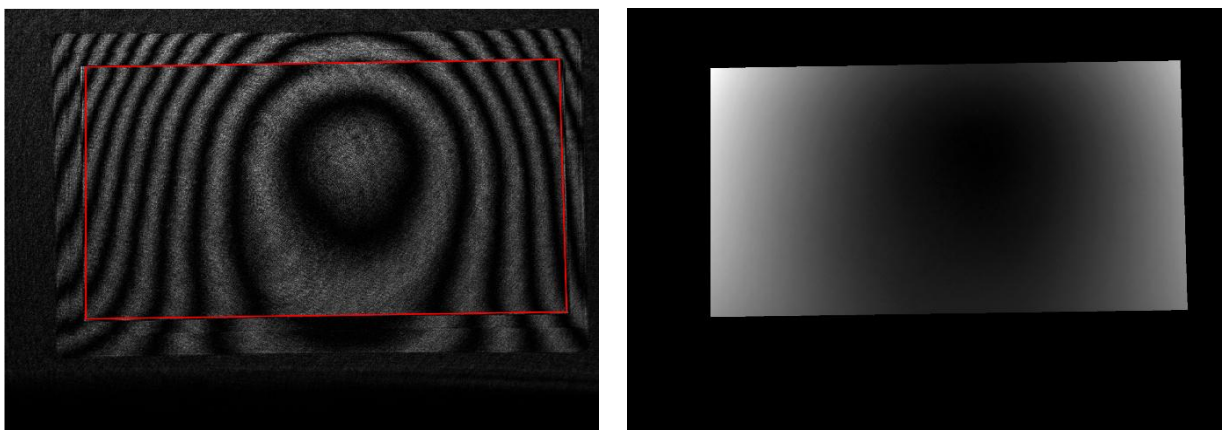

*Figure 3 Interferogram recorded at gray value 0 (left) and reconstructed surface from phase shift measurement (right).*

To identify possible phase errors, Zernike polynomials up to the order of 6 are fitted to and subtracted from the reconstructed surface shown in *Figure 4*. The residual height data resulting from that step is shown on the left-hand side of *Figure 4*. In addition, the right-hand side of *Figure 4* shows the Gamma value from the phase shift algorithm, which is a normalized measure of the fringe contrast. Both datasets, especially the fringe contrast, show that there are minor deviations from the perfect phase

6 | P a g e

step values, but with height amplitudes in the range of ± 30 nanometers, mostly dominated by higher frequency artefacts caused by dust particles, this can be considered a decent result for a first simple approach [22].

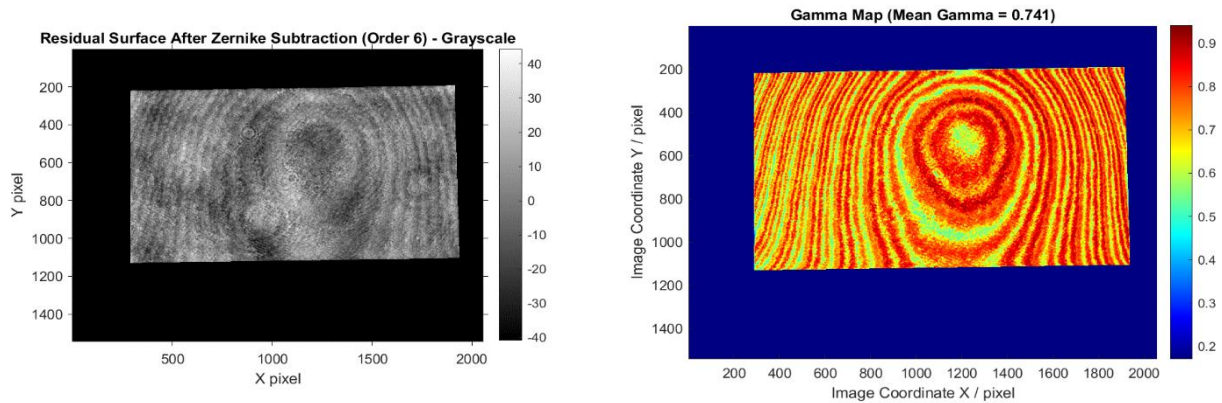

*Figure 4 Difference to a 6th order Zernike polynomial (left) and normalized fringe contrast 'Gamma' (right).*

**Section 6: Wavefront Compensation**

Based on the phase shift measurement, a wave front compensation has been performed. The inverse of the wrapped phase, computed as 2π minus the measured wrapped phase, has been applied to the LCOS display. The demanding part here is that a transformation from the image space of the camera to the image space of the LCOS display has to be determined. As can be seen from **Figure 3**, the active LCOS area appears slightly tilted and even slightly curved within the camera image. In a first simple approach, the curvature of the edges is neglected and a linear projective transformation is chosen. A transformation is analytically done with a MATLAB script that maps the corner coordinates of the LCOS as recorded by the camera to the corners of the LCOS pixel matrix. The image coordinates have been measured manually and the LCOS coordinates correspond to the corner coordinates is generated to form a Full HD image. The computed transformation of all pixel coordinates of the LCOS is mapped analytically to the corresponding camera image coordinates. The image coordinates are used to interpolate the required phase values from a phase measurement performed on the uncropped images.

As mentioned above, the (now transformed) measured values of the wrapped phase have been inverted by subtracting them from 2π. In the next step, the inverted phase values are mapped to corresponding gray values for driving the LCOS. This is done by applying the inverse of the linear function that has been used for the calculation of gray values that deliver the required phase shifts for a four-step measurement. With above assumptions the inverted phase is therefore divided by 2π and then multiplied with the reference gray value of 240. The matrix obtained from that step is rounded to integer values and then saved as 8-bit gray image. This image is then loaded to the Holoeye Pattern Generator App and applied to the LCOS display.

*Figure 5* shows the results obtained from the described workflow. The left-hand side of *Figure 5* i.e., *(a)*, *(b)*, *(c)* and *(d)* show the interferograms recorded with a global phase shift of 0, 60, 120 and 180 (which means the LCOS is driven with gray values of 0, 60, 120 and 180 over the whole area) and the right-hand side of *Figure 5* i.e., *(e)*, *(f)*, *(g)* and *(h)* show the interferograms with the inverse wrapped

# Phase shifting Interferometry and Wavefront Compensation using Liquid Crystal on Silicon (LCOS) Spatial Light Modulator (SLM)

**Indrani Bhattacharya[1], Marcus Petz[2], Rainer Tutsch[3]**

phase applied to the LCOS. As can be seen, due to the wavefront compensation, the intensity (and thus the phase) appears quite uniform within the active area of the LCOS. The thin lines resembling the interference pattern are the zones within the LCOS where a phase step between 0 and 2π occurs.

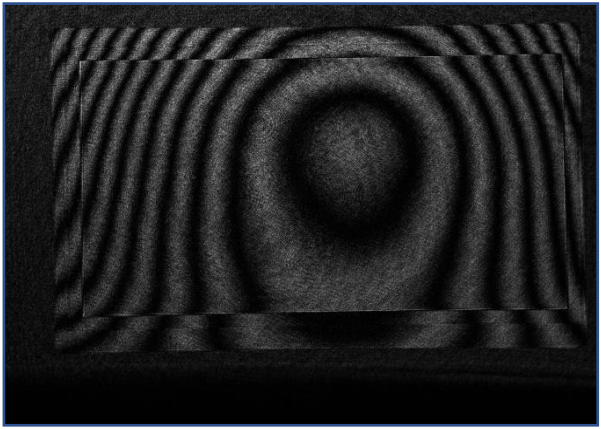

*(a) Gray value '0'*

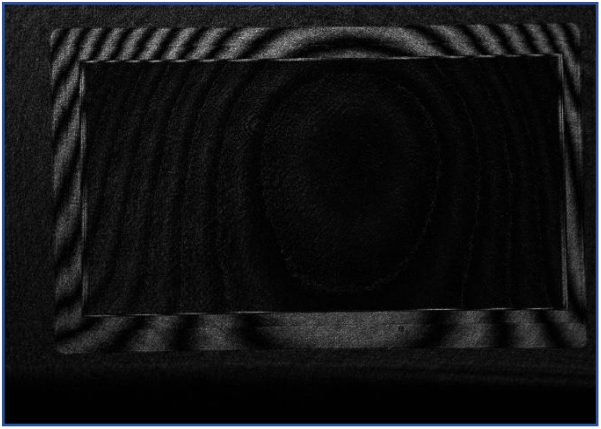

*(e )*

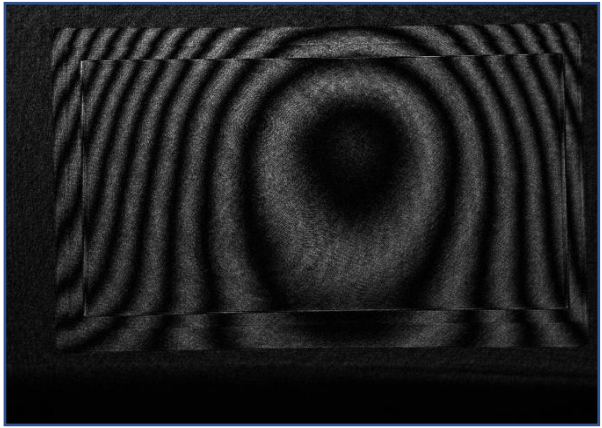

*(b) Gray value '60'*

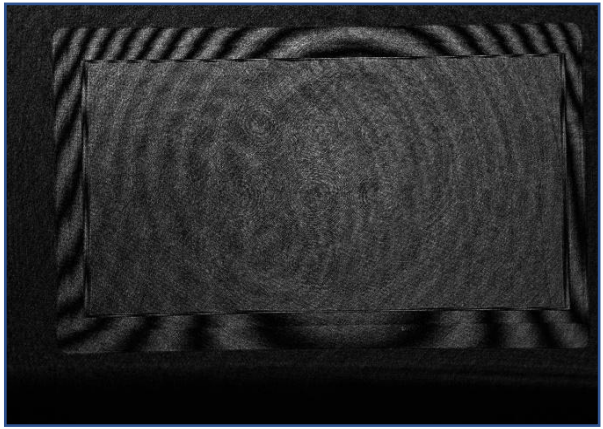

*(f)*

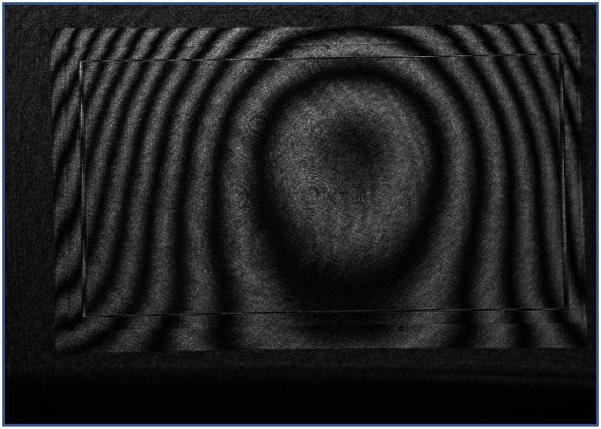

*(c ) Gray value '120'*

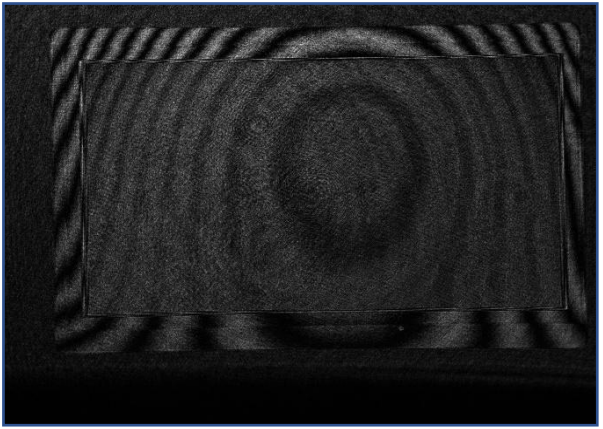

*(g)*

**Phase shifting Interferometry and Wavefront Compensation using Liquid Crystal on Silicon (LCOS) Spatial Light Modulator (SLM)**
**Indrani Bhattacharya[1], Marcus Petz[2], Rainer Tutsch[3]**

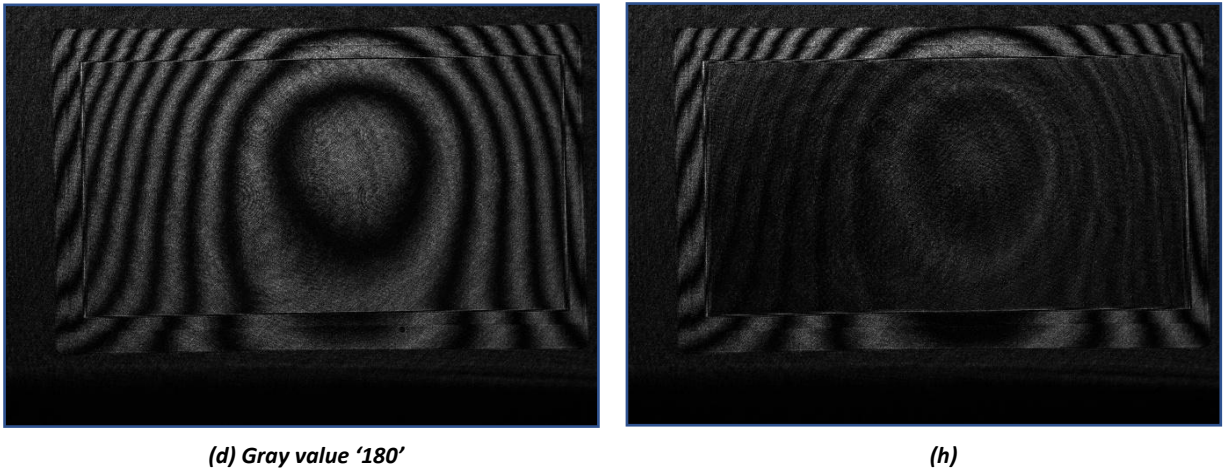

*(d) Gray value '180'*                  *(h)*

*Figure 5: (a),(b), (c) and (d) Interferograms recorded at gray values 0, 60, 120, 180 without wavefront compensation (left) and (e), (f), (g) and (h) Interferograms recorded with (simple proof of concept) wavefront compensation determined from a four-step phase shift measurement (right).*

From *Figure 5 (e)*, *(f)*, *(g)* and *(h)* the shape of the compensated wavefront can be calculated using the four-step algorithm. It should be nearly a plane wave front. *Figure 6 (a)* and *(b)* show the reconstructed wavefront surface before and after compensation.

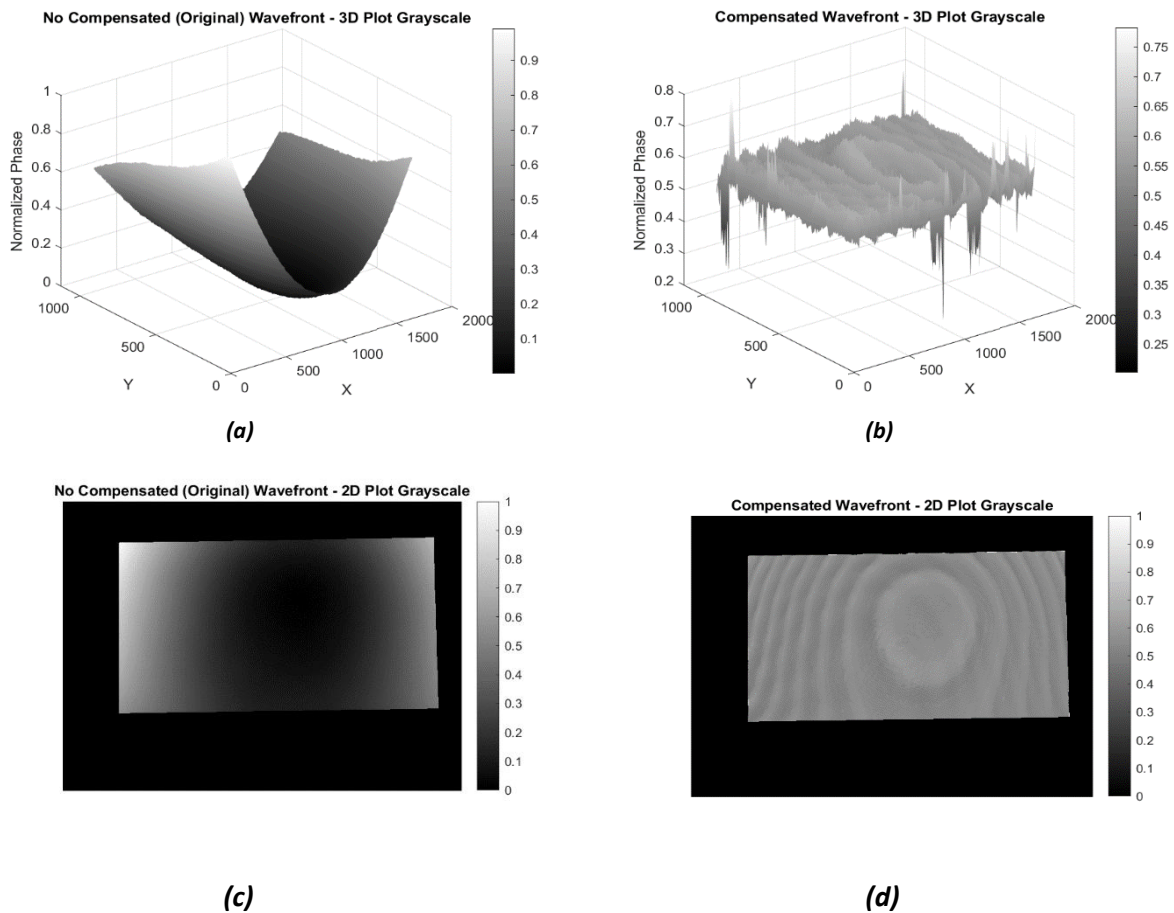

*(a)*                  *(b)*

*(c)*                  *(d)*

*Figure 6 (a)*, *(b), (c) and (d)* show the 3D and 2D plots of the reconstructed wavefronts using four-step algorithm before and after compensation.

9 | P a g e

**Section 7: Conclusion and Discussion**

A simple yet novel technique for interferometric surface measurement using LCOS SLM in Twyman Green Configuration is reported. Even though the tested approach is quite simple, the result shows no significant mapping errors.
In addition, it is important to mention that between the recording of the four-step measurement, which delivers the data used for the wavefront compensation, and the recording of the compensation itself has been done for a time difference of approximately 2 hours, during which the setup might have been influenced by the environment.

**Section 8: Further scope of Research**

In the phase shift algorithm discussed here, if the actual phase shift differs from the assumed value, errors may be introduced into the reconstruction of wavefronts. Non-uniformities of the phase-shift across the pupil caused by a linear phase shift in a converging beam may be another source of error. Carre algorithm may be tried to get a better result in this case.

*Acknowledgement:* Indrani Bhattacharya[1] expresses sincere thanks and heartfelt gratitude to Prof. Rainer Tutsch for inviting and providing scope for an insightful and collaborative research in the field of 'LCOS SLM Interferometry with an application in Optical Metrology', in the Institute of Production Metrology (IPROM), *Techniche Universitat, Braunschweig, Germany.* She expresses sincere thanks and heartfelt gratitude to Dr. Marcus Petz for invaluable guidance and suggestions to pursue this research work.
She also expresses sincere thanks and heartfelt gratitude to Ms. Annette Budin of IPROM to provide immense administrative help and guidance while pursuing the research work in the Institute.

**Phase shifting Interferometry and Wavefront Compensation using Liquid Crystal on Silicon (LCOS) Spatial Light Modulator (SLM)**
**Indrani Bhattacharya[1], Marcus Petz[2], Rainer Tutsch[3]**